\documentstyle[12pt,epsf]{article}
\renewcommand{\bar}[1]{\overline{#1}}

\begin{document}

\begin{flushright}
\end{flushright}

\bigskip\bigskip
\centerline{\large \bf Asymmetry of strange sea in nucleons}

\vspace{18pt}

\centerline{Xue-Qian Li$^{1,2}$, Xiao-Bing Zhang$^{1,2}$, and Bo-Qiang Ma$^{1,3}$}

\vspace{0.3cm}
\centerline{$^{1}$CCAST (The World Laboratory), P.O. Box 8730, Beijing, 100080}

\centerline{$^{2}$Department of Physics, Nankai University, Tianjin, 300071, China}

\centerline{$^{3}$Department of Physics, Peking University, Beijing, 100871,
China\footnote{Mailing address.}}

\vspace{1cm}


\vspace{10pt}
\begin{center} {\large \bf Abstract} \end{center}

Based on the finite-temperature field theory, we evaluate
the medium effects in nucleon
which can induce an asymmetry between quarks and antiquarks of the strange sea.
The short-distance effects determined by the weak interaction
can give rise to $\delta m\equiv \Delta m_s-\Delta m_{\bar s}$
where $\Delta m_{s(\bar s)}$ is the medium-induced mass of strange
quark by a few KeV at most, but the long-distance effects
by strong interaction are
sizable. Our
numerical results show that there exists an obvious
mass difference between strange
and anti-strange quarks, as large as $10-100$ MeV.

\vfill

\centerline{PACS numbers: 11.10.Wx, 11.30.Na, 12.40.-y, 14.20.Dh}

\vfill
\centerline{To be published in Phys.~Rev.~D.}
\vfill
\newpage


\section{Introduction and Motivation}

The existence of the Dirac sea is always an interesting topic
which all theoretists and experimentalists of high energy physics
are intensively pursuing, and the strange content of the nucleon
sea is of particular interest for attention. Ji and Tang \cite{Ji}
suggested that if a small locality of strange sea in nucleon is
confirmed, some phenomenological consequences can be resulted in.
The CCFR data \cite{CCFR} indicate that $s(x)/\bar s(x)\sim
(1-x)^{-0.46\pm 0.87}$. Assuming an asymmetry between $s$ and
$\bar s$, Ji and Tang analyzed the CCFR data and concluded that
$m_s=260\pm 70$ MeV and $m_{\bar s}=220\pm 70$ MeV \cite{Ji}. So
if only considering the central values, $\delta m\equiv
m_s-m_{\bar s}\sim 40$ MeV, which implies a quark-antiquark
asymmetry. However, one may also alternatively conclude that the
data are consistent with no asymmetry within err-bars
\cite{Ji,CCFR}.

In the framework of the Standard Model $SU(3)_c\otimes SU(2)_L\otimes U(1)_Y$,
we would like to look for some possible mechanisms which can induce the
asymmetry between quarks and antiquarks.

The self-energy of
strange quark and antiquark $\Sigma_{s(\bar s)}=\Delta m_{s(\bar
s)}$ occurs via loops where various interactions contribute to
$\Sigma_{s(\bar s)}$ through the effective vertices.
Obviously, the QCD interaction cannot distinguish between $s$ and $\bar s$,
neither the weak interaction alone in fact.
Practical calculations of the
self-energy also shows that $\Delta m_s=\Delta m_{\bar s}$.
In fact, because of the CPT theorem, $s$ and
$\bar s$ must be of exactly the same mass.
So we would ask
ourselves what can make an asymmetry between $s$ and $\bar s$ which are supposed
to be the sea quark-antiquark in nucleons. We can immediately find that in nucleons
there are asymmetric quarks $u$ and $d$, namely
the composition of $u$-$\bar u$
and $d$-$\bar d$ quark-antiqaurk is asymmetric.
Here $u$-$\bar u$, $d$-$\bar d$
include valence and sea portions of corresponding flavors.
In proton, there are two valence
$u$-quarks, but one valence $d$-quark, while in neutron, one $u$, but two
$d$'s. This asymmetry, as we show below, can stand as a medium effect which
results in an asymmetry $\Delta m_s\neq\Delta m_{\bar s}$ for strange sea
quarks.

As discussed above, if we evaluate the self-energy $\Delta m_s$ and $\Delta
m_{\bar s}$ in vacuum, the CPT theorem demands $\Delta m_s\equiv \Delta
m_{\bar s}$. However, when we evaluate them in an asymmetric environment of
nucleons, an asymmetry $\Delta^{M} m_s\neq \Delta^{M} m_{\bar s}$ where
the superscript $M$ denotes the medium effects, can be expected. In
other words, we suggest that the asymmetry of the $u$ and $d$ quark composition
in nucleons leads to an asymmetry of the strange sea.

There exist both short-distance and long-distance medium effects.
The short-distance effects occur at quark-gauge boson level,
namely a self-energy loop including a quark-fermion line and a
W-boson line or a tadpole loop (see below for details). The
contributions of $u$ and $d$-types of quark-antiquark to the
asymmetry realize through the Kabayashi-Maskawa-Cabibbo mixing.
Because of the small parton mass, the Higgs contributions can be
neglected. By contrary, the long-distance effects are caused by
loops which include a quark-fermion line and a meson (Kaon in our
case) line.

In fact, Brodsky and one of us proposed a meson-baryon resonance
mechanism and they suggested that the sea quark-antiquark
asymmetries are generated by a light-cone model of energetically
favored meson-baryon fluctuations \cite{Brodsky}.
It was first observed
by Signal and Thomas \cite{Sig87} that
the meson-cloud model of nucleon can introduce a mechanism
for the strange-antistrange asymmetry in the nucleon sea,
though their formalism is not consistent in treating the antiquark
distributions in a strict sense \cite{add}. An $s$-$\bar{s}$ asymmetry
was also predicted by Burkardt and
Warr \cite{Bur92} from the chiral Gross-Neveu model
at large $N_c$ in the LC formalism.
Our physics picture is
similar and the method is parallel to
theirs, while all the calculations are done
based on the finite temperature field theory.

\section{The Formulation}

 We are going to employ the familiar formulation of the Quantum Field
Theory at finite temperature and density. As well-known,
the thermal propagator of quarks can be written as \cite{ftft}
\begin{equation}
\label{pro}
i S_q(k)=\frac{i (\rlap /k+m_q)}{k^2-m_q^2}-2\pi(\rlap /k+m_q)\delta(k^2-m_q^2)f_F(k\cdot u),
\end{equation}
where $u_\mu$ is the four-vector for the medium and $f_F$ denotes
the Fermi-Dirac distribution function
\begin{equation}\label{ff}
f_F(x)={\theta(x)\over e^{\beta(x-\mu)}+1}+{\theta(-x)\over e^{-\beta (x-\mu)}
+1},
\end{equation}
and $\beta=1/kT$, $\mu$ is the chemical potential.
We notice that the first term of Eq.~(\ref{pro}) is just
the quark propagator in
the vacuum. Its contribution to $\Sigma_1$ is of no importance to
us because this is related to the wave-function renormalization of
the quark in the vacuum. We need to focus on the medium effect,
which comes from the second term of Eq.~(\ref{pro}).

It is experimentally confirmed that the total light quark number
in the nucleon is 3. If we omitted the small
mass differences of the light quarks ($u$ and $d$ types,
explicitly), the quark density in the nucleon is
\begin{equation}
\label{nq}
n_q-n_{\bar q}=\int {d^3k\over (2\pi)^3}[f_F(\omega_k)
-f_{F}(-\omega_k)]={3\over V_{eff}},    \, \,\,  ( q=u,d ).
\end{equation}
In this expression $\omega_k=\sqrt{{\bf{k}}^2+m_q^2}$ is
the energy of the light quark
and $V_{eff}$ is the effective nucleon volume where $q (\bar{q})$ resides.
Here we have ignored the possible sea quark asymmetry for light
quarks of $u$ and $d$-flavors \cite{Kum}.
For up and down flavors, we have
\begin{equation}
n_u-n_{\bar u}={2\over V_{eff}}, \,\,\,\, n_d-n_{\bar d}={1\over V_{eff}},
\end{equation}
in proton while
\begin{equation}
n_u-n_{\bar u}={1\over V_{eff}}, \,\,\,\, n_d-n_{\bar d}={2\over V_{eff}},
\end{equation}
in neutron.

\subsection{The short-distance contribution}

The corresponding Feynman diagrams are shown in Fig.1 (a) and (b).

\renewcommand{\thefigure}{1~(a)}
\setcounter{figure}{0}
\begin{figure}[htb]
\begin{center}
\leavevmode {\epsfysize=3.5cm \epsffile{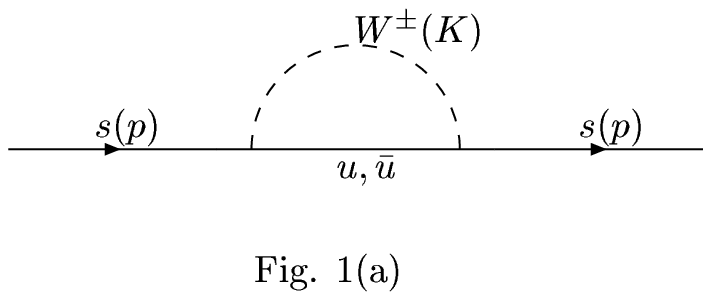}}
\end{center}
\caption[*]{\baselineskip 13pt
The self-energy $\Sigma^s_1$  for  strange quark
where the exchanged bosons are either W-boson or kaon
corresponding to short- and long-distance effects
respectively. }\label{strf1a}
\end{figure}

\renewcommand{\thefigure}{1~(b)}
\setcounter{figure}{0}
\begin{figure}[htb]
\begin{center}
\leavevmode {\epsfysize=6cm \epsffile{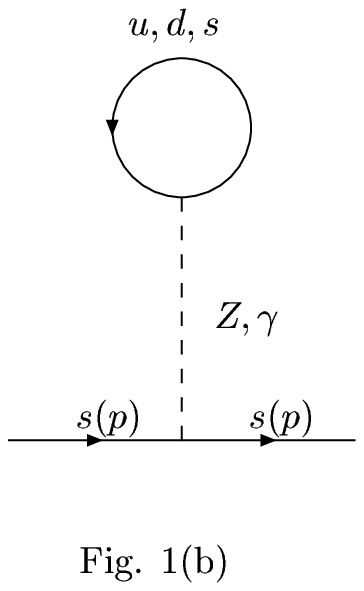}}
\end{center}
\caption[*]{\baselineskip 13pt
The tadpole diagram which contributes to the
self-energy $\Sigma^s_2$ of strange quark. }\label{strf1b}
\end{figure}

The two contributions  to the self-energy of $s$-quark ($\bar s$)
(a) and (b) are due to the charged current ($W^{\pm}$) and
neutral current (including the Z and
$\gamma$) respectively, the later is usually called as the tadpole-diagram \cite{Pal}.
The total contribution is in analog to that given in \cite{Pal}, the mere
difference is that  for
neutrino only weak neutral current plays a role while for quarks, the
electromagnetic interaction also needs to be involved.

The contribution due to the charged current is
\begin{equation}
\Sigma^{s}_1=\sqrt 2G_F\gamma^0L\sin^2\theta_C(n_u-n_{\bar u}),
\end{equation}
where $G_F$ is the Fermi coupling constant, $\theta_C$ is the Cabibbo
angle.
The contribution due to the weak neutral current is
\begin{equation}
\Sigma^{s}_2=3\sqrt 2G_F(-1+{4\over 3}Q^{(s)}\sin^2\theta_{_W})\cdot
\sum_f(T_3^{(f)}-2Q^{(f)}\sin^2\theta_{_W})(n_f-n_{\bar f}),
\end{equation}
where $Q^{(f)}$ refers to the charge of corresponding quark ($u$, $d$, $s$).
Pal and Pham
pointed that the axial part of the neutral current does not
contribute \cite{Pal}.

For the electromagnetic current the situation is different.
The exchanged photon connecting the $s$-quark line (or $\bar s$-quark) and the
closed loop (tadpole) possesses zero energy-momentum and its propagator
$${1\over q^2+i\epsilon}$$
results in an infrared divergence.
In the regular field theory, it does not bring up any problem
because  the integration over the inner momentum of the loop is
exactly zero. In the case of the gluon infrared
divergence, because of the non-Abelian Yang-Mills properties,
there can be a tachonic gluon mass \cite{Huber} which can serve as
an infrared cut-off. Unlikely, photon as the gauge boson of the
$U_{em}(1)$ group
cannot obtain an effective mass. It also means that there is no
any connection between the closed fermion-loop and the $s$-quark
line, thus the two parts are actually disconnected and
such electromagnetic tadpole should not be
included in the phenomenological calculations even though it has
a superficial infrared divergence.
Thus we drop out the electromagnetic tadpole in our later calculations.

As for the strange antiquark, one can obtain the corresponding
self-energy by changing the direction of propagation. Obviously if
ignore the small mass difference of $u$ and $d$-quarks, we have
$$\Sigma_1^{\bar s}=-\Sigma_1^{s},\;\;\;{\rm and}\;\;\;
\Sigma_2^{\bar s}=-\Sigma_2^{s},$$ and the mass difference between
the  strange and antistrange quarks is
\begin{equation}
\delta m\equiv 2(\Sigma_1^s+\Sigma_2^s),
\end{equation}
due to the short-distance interactions.

\subsection{The long-distance effects}

Up to now the dynamical picture of the sea quark interaction in
QCD is not definitively understood. In our case of interest, the
main contribution of the strong interaction will be attributed to
the low-energy effective coupling between the internal Goldstone
bosons and quarks. According to the common knowledge of low-energy
effective strong interaction, the strange quark is generated from
dissociation of nucleons into hyperons plus kaons.
In this picture the $s(\bar
s)-$quark in the sea interacts with the light quark and the kaon
meson essentially. This is in analog to that considered in
Refs.~\cite{Brodsky,Sig87,Bur92} and by Kogan and Shifman who
introduced such effects for weak radiative decay \cite{Kogan}, but
we estimate these effects with a quark-meson interaction instead
of a baryon-meson interaction \cite{Yan1}.

In the calculations, we need an effective vertex for $\bar
sqM$ where $q$ can be either $u$ or $d$-quarks and $M$ is a
pseudoscalar or vector meson. Here we only retain the lowest
lying meson states such as $\pi,K,\rho$ etc. The effective
chiral Lagrangian for the interaction between quarks and
mesons has been derived by many authors \cite{Georgi,Yan}.
For completeness, we present the well-established form of
the chiral Lagrangian as \cite{Yan}
\begin{equation}
L_x=i\bar q(\rlap /{\partial}+\rlap /\Gamma +g_A\rlap
/\Delta \gamma_5-i\rlap /{\nabla} ) q-m\bar qq+{\rm meson\; part}.
\end{equation}

It is noted that in general, the chiral Lagrangian only applies to the interactions 
between constituent quarks and mesons, 
on other side the sea quark picture is valid for current quarks (or partons)
\cite{Zhi}. Here we borrow the chiral Lagrangian picture just because the fundamental 
forms of interactions of either partons or constituent quarks with mesons
are universal.
The key point is the coupling constants, they may be different in the two pictures. 
However, we can assume that they do not deviate by orders from each other. 
Therefore we can use the coupling
constants for constituent quarks in our estimation of 
the order of magnitude. Probably, 
the obtained results and new data 
would offer us an opportunity to determine the effective coupling 
between parton-quarks and mesons. 
Anyway, we would point out that the numerical results obtained 
in this work may have a relative large error of about 10 MeV, as we will see
in the section of numerical calculations.

In this expression we omit the irrelevant part which only
contains mesons. Here $\bar q=(\bar q_u,\bar q_d,\bar q_s)$
and
\begin{equation}
V_{\mu}(x)={\bf \lambda}\cdot {\bf V}_{\mu}=\sqrt 2
\left( \begin{array}{ccc}
{\rho^0_{\mu}\over\sqrt 2}+{\omega_{\mu}\over\sqrt 2} &
\rho^+_{\mu} & K^{*+}_{\mu} \\
\rho_{\mu}^- & -{\rho^0_{\mu}\over\sqrt 2}+
{\omega_{\mu}\over\sqrt 2} & K^{*0}_{\mu} \\
K^{*-}_{\mu} & \bar K^{*0}_{\mu} & \phi_{\mu}
\end{array}\right).
\end{equation}
$\Delta_{\mu}$ and $\Gamma_{\mu}$ are defined as
\begin{eqnarray}
\Delta_{\mu} &=& {1\over 2}
(\xi^{\dag}(\partial_{\mu}-ir_{\mu})\xi-
\xi(\partial_{\mu}-il_{\mu})\xi^{\dag}), \nonumber\\
\Gamma_{\mu} &=& {1\over 2}
(\xi^{\dag}(\partial_{\mu}-ir_{\mu})\xi+
\xi(\partial_{\mu}-il_{\mu})\xi^{\dag}),
\end{eqnarray}
with $$\xi=exp(i\lambda^a\Phi^a(x)/2f),$$
where the Goldstone bosons $\Phi^a$ are the pseudoscalar
mesons in the SU(3) octet and $f$ is the decay constant.

From the chiral effective Lagrangian,
the basic effective vertex is a
pure-derivative axial vector (chiral-symmetric) coupling  as
$f_{kqs}{\bar \psi} {\gamma_5}{\gamma_\mu} {\partial_\mu} K \psi$.
In the last part of this work, we will give more
discussions on this issue.
The quantum correction at one-loop level provides a self-energy
$\Sigma_3^s$ for the strange quark shown in Fig.~1(a) (where we only need
to replace
the W-boson line by a kaon-line).
Here we neglect the higher loop contribution which may be induced
by the higher order graphs in the chiral Lagrangian. In general,
for such an estimation, the effective coupling by itself may contain some
higher loop contributions, thus their effects do not influence
the qualitative conclusion although the quantitative result may
change \cite{Yan}.
So we can write the
amplitude due to the long-distance effective interaction as
\begin{equation}
\label{1}
-i \Sigma_3^s=i {f^2_{kqs}}\int \frac{d^4 k}{{(2\pi)}^4}
{\gamma_5}{\gamma_\mu}iS_q(k){\gamma_5}{\gamma_\mu}
\frac{{(p-k)}^2}{{(p-k)}^2-M^2_K},
\end{equation}
where $M_K$ is the mass of kaons.

In the rest frame of the medium ( $u_\mu=(1,\mathbf{0})$ ) we use
\begin{equation}
\delta (k^2-m_q^2) = \frac{1}{2\omega_k}[\delta(k_0-\omega_k)
+\delta(k_0+\omega_k)] ,
\end{equation}
where $\omega_k=\sqrt{\mathbf{k}^2+m_q^2}$ is the energy of the light
quark. So the long-distance medium correction to the  mass of strange
quark can be evaluated and we obtain
\begin{equation}
\label{m}
\Sigma_3^s = \gamma_0 \frac{f^2_{kqs}}{2}{\int}\frac{d^3\mathbf{k}}{{(2\pi)}^3}
 [ \frac{m_s^2-2m_s \omega_k}{m_s^2-2m_s \omega_k-M_K^2 } f_F(\omega_k)
-\frac{m_s^2+2m_s \omega_k}{m_s^2+2m_s \omega_k-M_K^2} f_{F}(-\omega_k) ].
\end{equation}
After simple manipulations, it becomes
\begin{eqnarray} \label{mm}
\Sigma_3^s &=& \gamma_0\frac{f^2_{kqs}}{2} [ (n_q-{n_{\bar q}})
+ \int \frac{d^3\mathbf{k}}{{(2\pi)}^3}
\frac{M_K^2}{m_s^2-2m_s \omega_k-M_K^2} f_F(\omega_k) \nonumber \\
& &-\int \frac{d^3\mathbf{k}}{{(2\pi)}^3}
\frac{M_K^2}{m_s^2+2m_s \omega_k-M_K^2}f_{F}(-\omega_k)
].
\end{eqnarray}
In order to
avoid the pole in the second term  of Eq.~(\ref{m}), we use the familar Breit-Wigner
formulation. Then we give $\Delta m_s$ contributed by the long-distance effects as
\begin{eqnarray}
\label{pra}
\Delta m_s &=& \frac{f^2_{kqs}}{2}{\int}\frac{d^3\mathbf{k}}{{(2\pi)}^3}
 [ \frac{(m_s^2-2m_s \omega_k-M_K^2)(m_s^2-2m_s \omega_k)}
 {(m_s^2-2m_s \omega_k-M_K^2)^2+{(M_K \Gamma_K)}^2}
  f_F(\omega_k) \nonumber \\
& &-\frac{(m_s^2+2m_s \omega_k-M_K^2)(m_s^2+2m_s \omega_k)}
{(m_s^2+2m_s \omega_k-M_K^2)^2+{(M_K \Gamma_k)}^2}
 f_{F}(-\omega_k) ]         \nonumber \\
  &=& \frac{f^2_{kqs}}{2}{\int}\frac{{k^2}d {k}}{2\pi^2}
 [ \frac{(m_s^2-2m_s \omega_k-M_K^2)(m_s^2-2m_s \omega_k)}
 {(m_s^2-2m_s \omega_k-M_K^2)^2+{(M_K \Gamma_k)}^2}
  f_F(\omega_k) \nonumber \\
& &-\frac{(m_s^2+2m_s \omega_k-M_K^2)(m_s^2+2m_s \omega_k)}
{(m_s^2+2m_s \omega_k-M_K^2)^2+{(M_K \Gamma_k)}^2}
 f_{F}(-\omega_k) ],
\end{eqnarray}
where in the Breit-Wigner expression, we take the
usual approximation that
$\Gamma_K$ is the measured value of the lifetime of
$K^{\pm}$.

The statistical integral Eq.~(\ref{pra}), as a function of
temperature $T$ and chemical potential $\mu$, can be expressed in
terms of the quark density and therefore the nucleon
size ( c.f. Eq.~(\ref{nq}) )
at a given temperature. In the practical calculation,
we use the numerical integral method
to relate the mass correction $\Delta m_s$
with the effective nucleon radius $R$.
\\

\section{The Numerical Results}

\subsection{For short-distance effects}


Our numerical results show that
$$\delta m= 92\ {\rm eV} \sim 0.8\ {\rm KeV}, \;\;\;\; {\rm for\; proton},$$
$$\delta m=0.38\ {\rm KeV} \sim 3.0\ {\rm KeV}, \;\;\;\; {\rm for\; neutron},$$
in the range of the effective nucleon radius $R\approx0.5-1.0$ fm.
So we see that the short-distance interaction cannot result in a
large asymmtry in the strange sea.

\subsection{For the long-distance effects}

According to the picture of chiral field theory \cite{eich,cheng,song,szcz},
the effective pseudovector
coupling implies $f_{kqs}=\frac{g_A}{\sqrt 2 f}$, where the axial-vector coupling
$g_A=0.75$. The pion decay constant $f_{\pi}=93$ MeV, kaon decay constant
$f_{K}=130$ MeV, for our estimation, an approximate SU(3) symmetry
might be valid, so that $f$ can be taken as an average of
$f_{\pi}$ and $f_K$.
One can trust that the order of
the effective coupling at the vertices does not deviate too much
from this value.

In the chiral quark model from the framework of the standard chiral field theory
\cite{Georgi},
one usually takes the quarks as the constituent quarks \cite{eich,cheng2}. However,
there is also a suggestion \cite{weber} to consider the quarks in the chiral dynamics
as current
quarks from the successful description of the proton spin data \cite{cheng,song},
in consistent with our consideration to use the chiral Lagrangian picture 
for an effective description of current-quark and meson interaction.
Therefore we present our calculated results of $\Delta m_s$ for two cases:\\
(a) with the quark masses being the current quark
masses $m_s=150$~MeV and $m_u \approx m_d=6$ MeV; \\
(b) with the quark masses being the constituent quark masses
$m_s=500$~MeV and $m_u=m_d=350$~MeV.\\

We present the numerical results of $\Delta m_s$ as a function
of the effective nucleon radius
in Fig.~2. We find that the result depends on the value of
nucleon volume sensitively for small $R$-values, which correspond to the high density,
but becomes mild as $R$ turns larger.

\renewcommand{\thefigure}{2}
\setcounter{figure}{1}
\begin{figure}[htb]
\begin{center}
\leavevmode {\epsfysize=5cm \epsffile{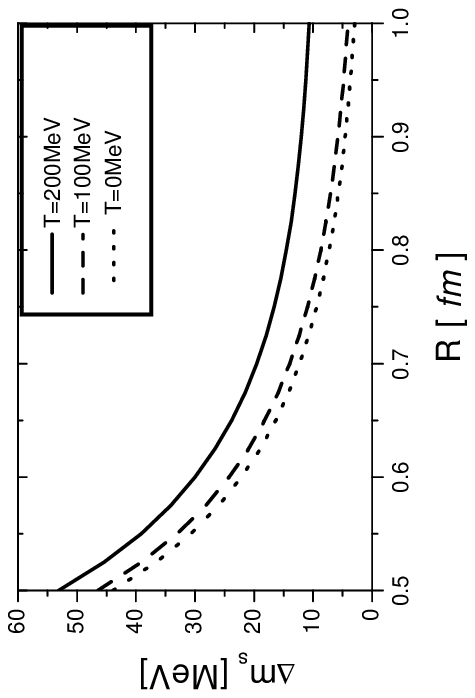}}
\end{center}
\caption[*]{\baselineskip 13pt
The medium correction $\Delta m_s$ for the strange quark mass vs
the effective nucleon radius $R$, in two cases with different quark masses:
(a) current quark masses $m_s=150$~MeV and $m_u \approx m_d=6$ MeV; 
(b) constituent quark masses
$m_s=500$~MeV and $m_u=m_d=350$~MeV. }\label{strf2}
\end{figure}

In Fig.~2, one can notice that
as $R$ is  about 0.5 fm, $\Delta m_s$ can be as large as 50 MeV which results in
$\delta m\equiv 2\Delta m_s\simeq100$ MeV for the current quark mass case, 
and even larger for the constituent quark mass case.
But for the situation corresponding to a normal nucleon case, $\delta m$
might be only of the order of around a few 10 MeV, which is
within the uncertainties as estimated in Ref.~\cite{Ji}.

We notice that at high temperature with ordinary density of
nuclear matter, or at high density with ordinary temperature, the
baryons and mesons may undergo a phase transition  to the
quark-gluon-plasma with quarks and gluons as the basic degrees of
freedom, and spontaneous chiral symmetry may be restored. Thus the
effective $s$-kaon-$u$ coupling $f_{kqs}$ and the pion or kaon
decay constant $f$ may decrease with increasing temperature and
density, and the quark-antiquark asymmetry may also decrease with
increasing temperature and density. However, here we consider the
nucleon case, with $T$ refers to the temperature of normal nuclear
matter in the chiral dynamics, in a range of about 100 MeV to a
few hundreds MeV, which is lower than chiral symmetry restoration
scale of about 1 GeV, and also with the density is not high for a
nucleon as ordinary nuclear matter. Therefore we neglect the
dependence of the coupling and decay constants on temperature and
density in the present study. But it should be clear that the
trend as shown in Fig.~2, where the strange quark-antiquark mass
difference $\delta m_s$ increases with increasing temperature and
density, will be changed if the chiral symmetry restoration is
considered. Thus the quark-antiquark asymmetry caused by the
long-distance effects should not be very larger than what we
estimated above, if such asymmetry can be large.

\section{Discussions and Summary}

The strange asymmetry discussed above in fact may be caused by an
asymmetry of the light quark contents in nucleons, but not the
interaction itself. We show that $u(\bar u), d(\bar d)$ quarks in
nucleons can play a role as the medium which results in an
asymmetry of the self-energy of $s$ and $\bar s$. The resultant
values for $s$ and $\bar s$ have opposite signs which  induce a
net mass difference between $s$ and $\bar s$. The nucleon
structure determines $\delta m=m_s- m_{\bar s}>0$.

There exist both short-distance and long-distance effects which
are mediated by W (Z) bosons and K-mesons respectively. Because
the gauge bosons W and Z are very heavy, the net effects are much
suppressed and their contribution to $\delta m$ can only be a few
KeV. By contrast, kaon is much lighter and moreover, the effective
interaction $\bar qqP$ where P denotes a meson with appropriate
quantum number is due to non-perturbative QCD which is strong
interaction, thus the effective coupling is much larger than that
of weak interaction. This obvious enhancement gives a value of
10$\sim$100 MeV to $\delta m$, which is consistent with the
estimate \cite{Ji} required to fit the experimental data within
err-bars.

In our approach, we only keep the leading order in the
effective chiral Lagrangian and leave the coupling constant
as a parameter with an SU(3) symmetry. If we completely apply the chiral
effective Lagrangian, the coupling constant is fully determined.
In other side,
the original formulation of the effective chiral Lagrangian
is for the constituent quark whereas here we take the
parton picture instead, thus the coupling constant might somewhat
deviate from that in the effective chiral Lagrangian. In this work
we take the value of the coupling constant according to
the Lagrangian but with a factor $g_A$ as $f_{kqs}={g_A\over \sqrt
2 f}$. This factor $g_A$ partly includes the nucleon structure
effects and compensates the errors as applying the chiral
effective Lagrangian to the parton picture.
We can expect that this choice does  not much deviate from
reality while using the Lagrangian for the parton picture.
Certainly, such approximation may bring up errors, so that
we cannot give precise predictions on the mass difference
of $\bar s$ and $s$, but one can be convinced that
the qualitative conclusion can be
made and the order of magnitude is close to reality, because here we
apply all the established theories and models except the numerical
value of the coupling constant to make this evaluation.

In summary, we investigate the influence of the  environment on
the asymmetry of the strange sea based on the well-established
finite-temperature field theory. In our opinion, a natural
explanation for this asymmetry is due to the light quark sea in
nucleons which stands as a medium with certain asymmetries. There
exist always three ``net" light quarks which means an excess of
$q$ over $\bar q$ in nucleons. The important feature is attributed
to a non-zero chemical potential for the light quark (antiquark)
in this theoretical framework. As a consequence, we obtain an
obvious asymmetry of the strange sea ($\delta m\sim 10-100$ MeV),
although the magnitude of the asymmetry should be also constrained
by chiral symmetry restoration. Considering that the net strange
quark number is zero, it leads to $\mu_s\neq\mu_{\bar s}\neq 0$.
We expect that this point is helpful for understanding the
non-negligible strangeness content of the nucleon. On the other
hand, we suggest that the low-energy effective interaction
provides the main dynamical origin of the strange sea asymmetry.
In fact, many authors have emphasized that the light-quark sea
asymmetry can arise from effective interactions between the
internal Goldstone bosons and quarks. So the essence of our
approach is consistent with the treatments of
\cite{eich,cheng,song,szcz} in the dynamical sense.

\noindent
{\bf Acknowledgment:}

This work is partially supported by the National Natural Science
Foundation of China.\\

\vspace{1cm}

\end{document}